\documentclass[prb,twocolumn]{revtex4-2}
\usepackage{graphicx} 

\usepackage{amsmath,amsthm,amssymb,mathrsfs}
\usepackage{bm}
\usepackage{color}

\begin{document}

\title{Chaotic magnetization dynamics driven by feedback magnetic field}

\author{Tomohiro Taniguchi}
\email{tomohiro-taniguchi@aist.go.jp}
\affiliation{
National Institute of Advanced Industrial Science and Technology (AIST), Research Center for Emerging Computing Technologies, Tsukuba, Ibaraki 305-8568, Japan. 
}

\date{\today} 
\begin{abstract}
{
An excitation of highly nonlinear, complex magnetization dynamics in a ferromagnet, for example chaos, is a new research target in spintronics. 
This technology is applied to practical applications such as random number generator and information processing systems. 
One way to induce complex dynamics is applying feedback effect to the ferromagnet. 
The role of the feedback electric current on the magnetization dynamics was studied in the past.  
However, there is another way to apply feedback effect to the ferromagnet, namely feedback magnetic field. 
In this paper, we developed both numerical and theoretical analyses on the role of the feedback magnetic field causing complex magnetization dynamics. 
The numerical simulation indicates the change of the dynamical behavior from a simple oscillation with a unique frequency to complex dynamics such as amplitude modulation and chaos. 
The theoretical analyses on the equation of motion qualitatively explain several features found in the numerical simulations, exemplified as an appearance of multipeak structure in the Fourier spectra. 
The difference of the role of the feedback electric current and magnetic field is also revealed from the theoretical analyses. 
}
\end{abstract}

\maketitle


\section{Introduction}
\label{sec:Introduction}


Magnetization dynamics studied in magnetism and spintronics have mainly been focused on simple dynamics such as magnetization switching, ferromagnetic resonance, and microwave oscillation driven by magnetic field and/or electric current \cite{vonsovskii66,katine00,bertotti01,myers02,kiselev03,tulapurkar05,kubota05,diao05,tulapurkar05,kim06,houssameddine07,sankey08,kubota08,taniguchi08,okamoto12,taniguchi14,taniguchi16,lebrun17}. 
Recently, however, an excitation of complex dynamics, for example chaos, is studied from both experimental and theoretical perspectives  \cite{li06,yang07,watelot12,pylypovskyi13,devolder19,bondarenko19,montoya19,taniguchi19JMMM,yamaguchi19,taniguchi24} because of the possibilities in developing new practical applications, such as true-random-number generator and brain-inspired computing \cite{grollier16,grollier20}.  
One way to excite complex dynamics in physical, chemical, and biological systems is using feedback effects \cite{biswas18}. 
They are frequently observed in natural and artificial systems, such as electric circuit, population dynamics, and neural networks.
From the mathematical viewpoint, feedback effects significantly increase the number of the dynamical degrees of freedom of systems, or dimensions of phase space. 
Recall that the number of dynamical degrees of freedom strongly relates to dynamics possibly excited in a system. 
For example, chaos cannot be excited in a system described by a differential equation, whose dimension is less than or equal to $2$. 
This fact is known as the Poincar\'e-Bendixson theorem \cite{strogatz01,alligood97,ott02} and is one reason why it has been difficult to identify chaos in ferromagnets, where magnetization dynamics were well described by various simple models including only two dynamical variables, such as macrospin, vortex, spin wave, domain wall, and so on. 
Thus, several approaches to increase additional dynamical degree of freedom, such as vortex-core reversal \cite{pylypovskyi13,devolder19}, have been proposed \cite{li06,yang07,watelot12,bondarenko19,montoya19,taniguchi19JMMM,yamaguchi19,taniguchi24}. 
Adding feedback effects to ferromagnets is one of these approaches to overcome this issue and expect to excite rich dynamics which could appear in high-dimensional systems. 


The previous works on the feedback effect on spintronics devices mainly focus on using electric current as feedback signal \cite{khalsa15,tsunegi16,williame19,taniguchi19,williame20}. 
There is, however, another way to add the feedback effect to the magnetization dynamics, namely, converting the feedback signal into magnetic field \cite{kamimaki21,tsunegi23}. 
Although using magnetic field is not preferable for practical applications, there are several motivations to investigate the role of the feedback magnetic field on the magnetization dynamics. 
One is the fact that the torques due to the magnetic field and electric current have different roles on the dynamics. 
The former induces a sustainable oscillation of the magnetization while the latter induces non-conservative torque through spin-transfer effect \cite{slonczewski96,berger96}. 
The role of the feedback magnetic field on the dynamics however still remains unclear. 
Another motivation relates to an experimental viewpoint. 
The previous works have mainly focused on applying feedback current to spin-torque oscillators (STOs), and typical STOs consist of magnetic tunnel junctions (MTJs). 
MTJs include an insulating barrier such as MgO, and thus, there is a limitation on the magnitude of the feedback current to avoid electrostatic breakdown of the tunnel barrier. 
This was one reason why the past works dealt only in a weak feedback limit \cite{khalsa15,tsunegi16}. 
If we use, on the other hand, the feedback magnetic field, this issue will be circumvented \cite{kamimaki21,tsunegi23} because the feedback signal is not injected into MTJs directly. 
However, precise studies of theoretical analyses on the magnetization dynamics in the presence of feedback magnetic field have not been developed yet. 


In this paper, we study the role of feedback magnetic field on magnetic vortex dynamics using numerical simulation of the Thiele equation. 
Here, we evaluate both temporal dynamics of the vortex core and Fourier spectra and find an appearance of chaos in a large feedback limit. 
We also develop theoretical analyses on the feedback effect on the vortex oscillation. 
We derive equations which self-consistently determine the relationship between electric current, magnetic field strength, and delay time, and their solutions show qualitative agreement with the results of the numerical simulations. 
The difference of the role of the feedback electric current and magnetic field is also clarified by deriving a theoretical formula of a threshold current density. 


\section{System description}
\label{sec:System description}


\begin{figure}
\centerline{\includegraphics[width=1.0\columnwidth]{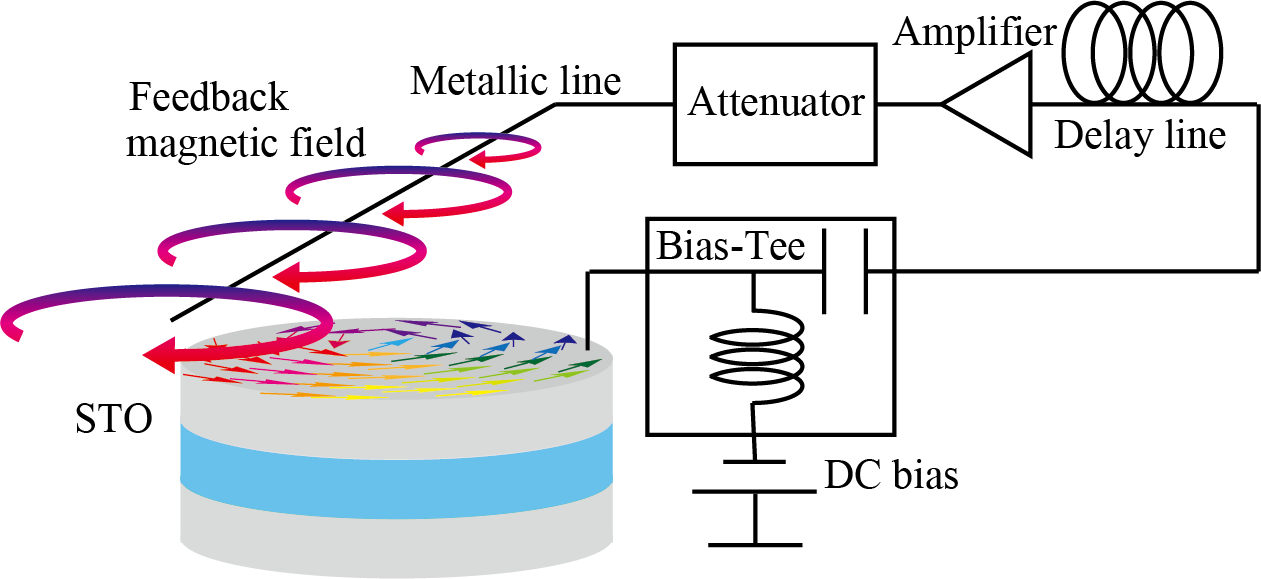}}
\caption{
            Schematic illustration of a vortex STO with a feedback circuit. 
            DC bias generates direct current, which induces vortex-core dynamics via spin-transfer torque effect. 
            Output signal can be separated from the input direct current by bias tee and is sent to a delayed feedback circuit with amplifier and/or attenuator. 
            Finally, the output signal flows in a metallic line as electric current and generates feedback magnetic field, which affects the vortex-core dynamics. 
         \vspace{-3ex}}
\label{fig:fig1}
\end{figure}


Figure \ref{fig:fig1} is a schematic illustration of a vortex STO with the feedback circuit \cite{kamimaki21}. 
Direct current is injected into the STO and excites an auto-oscillation of the vortex core. 
The dynamics modulates resistance of the STO due to the magnetoresistance effect, and therefore, the output signal oscillates,  which depends on the vortex-core position. 
The input direct current and the output signal can be separated by bias tee. 
The output signal is sent to a delayed-feedback circuit and passes through a metallic line, which is placed on the STO. 
The electric current flowing in the metallic line generates magnetic fields, whose magnitude reflects the vortex-core position with delay time $\tau$. 
The magnitude of the feedback magnetic field can also be tuned by an attenuator and/or amplifier. 
The dynamics of the vortex core in the presence of the feedback magnetic field is described by the following Thiele equation \cite{thiele73,guslienko06PRL,guslienko06,khvalkovskiy09,guslienko11,dussaux12,grimaldi14}, 
\begin{widetext}
\begin{equation}
\begin{split}
  &
  -G \mathbf{e}_{z}
  \times
  \dot{\mathbf{X}}
  -
  |D|
  \left(
    1
    +
    \xi 
    s^{2}
  \right)
  \dot{\mathbf{X}}
  -
  \kappa
  \left(
    1
    +
    \zeta 
    s^{2}
  \right)
  \mathbf{X}
  +
  a_{J} J p_{z}
  \mathbf{e}_{z}
  \times
  \mathbf{X}
\\
  &+
  c a_{J} J R_{0} 
  p_{x}
  \mathbf{e}_{x}
  -
  c b_{J} J R
  p_{x}
  \mathbf{e}_{y}
  +
  c \mu^{*}
  \mathbf{e}_{z}
  \times
  \mathbf{H}
  =
  \bm{0}, 
  \label{eq:Thiele}
\end{split}
\end{equation}
\end{widetext}
where $\mathbf{e}_{i}$ is the unit vector representing the $i$ direction, $\mathbf{X}=(X,Y,0)$ is the position vector of the vortex core, while $G=2\pi pML/\gamma$ and $D=-(2\pi\alpha ML/\gamma)[1-(1/2)\ln(R_{0}/R)]$ \cite{guslienko06} consist of the saturation magnetization $M$, the gyromagnetic ratio $\gamma$, the Gilbert damping constant $\alpha$, the thickness $L$, the disk radius $R$, and the core radius $R_{0}$. 
The dimensionless parameter $\xi$ was introduced to describe the nonlinear damping in a highly excited state \cite{dussaux12}. 
The magnetic potential energy $W$ is characterized by $\kappa$ and $\zeta$ via 
\begin{equation}
  W
  =
  \frac{\kappa}{2}
  \left(
    1
    +
    \frac{\zeta}{2}
    s^{2}
  \right)
  |\mathbf{X}|^{2}, 
\end{equation}
and $\kappa=(10/9)4\pi M^{2}L^{2}/R$ \cite{dussaux12}. 
The dimensionless parameter $\zeta$ was introduced to explain the linear dependence of the oscillation frequency on the current \cite{dussaux12}. 
We also introduce $s=|\mathbf{X}|/R=\sqrt{X^{2}+Y^{2}}/R$, which corresponds to the distance of the vortex-core position measured from the disk center. 
The spin-transfer torque strength with spin polarization $P$ is $a_{J}=\pi\hbar P/(2e)$ \cite{khvalkovskiy09,guslienko11}, while the strength of the field-like torque is given by $b_{J}$. 
The electric current density is denoted by $J$, where positive current corresponds to that flowing from the reference to the free layer. 
The unit vector pointing in the magnetization direction in the reference layer is assumed to lie in the $xz$-plane as $\mathbf{p}=(p_{x},0,p_{z})$. 
Accordingly, the output signal depends on the $y$ component $Y$ of the vortex-core position. 
The external magnetic field is denoted as $\mathbf{H}$, while we define $\mu^{*}$ as $\mu^{*}=\pi MLR$. 
Recall that the magnetic field is generated as the feedback effect and the output signal depends on $Y$. 
Therefore, $\mathbf{H}$ can be expressed as 
\begin{equation}
  \mathbf{H}(t)
  =
  h_{y}
  \frac{Y(t-\tau)}{R}
  \mathbf{e}_{y}, 
\end{equation}
where $h_{y}$ determines the magnitude of the feedback magnetic field, while $Y(t-\tau)/R$ is the normalized $y$ component of the vortex core position in the past with a delay time $\tau$. 
Here, we assume that the feedback magnetic field points to the $y$ direction. 
In the following, we express the magnitude of $h_{y}$ in terms of gain in the feedback circuit, which is frequently used in experiments \cite{kamimaki21,tsunegi18} (see also Appendix \ref{sec:AppendixA}). 
The values of the parameters are derived from typical experiments and summarized in Appendix \ref{sec:AppendixA}. 
We, however, inform here that the delay time $\tau$ is set to be $29$ ns \cite{tsunegi23}, which is sufficiently longer than the typical oscillation period of the vortex STO (about $3$ ns). 
This is because complex dynamics can be easily excited when the delay time is longer than typical time scale of systems \cite{biswas18}, while dynamics excited by a short-time delay effect were, for example, self-injection locking \cite{khalsa15,tsunegi16}.


\section{Numerical simulation}
\label{sec:Numerical simulation}


\begin{figure}
\centerline{\includegraphics[width=1.0\columnwidth]{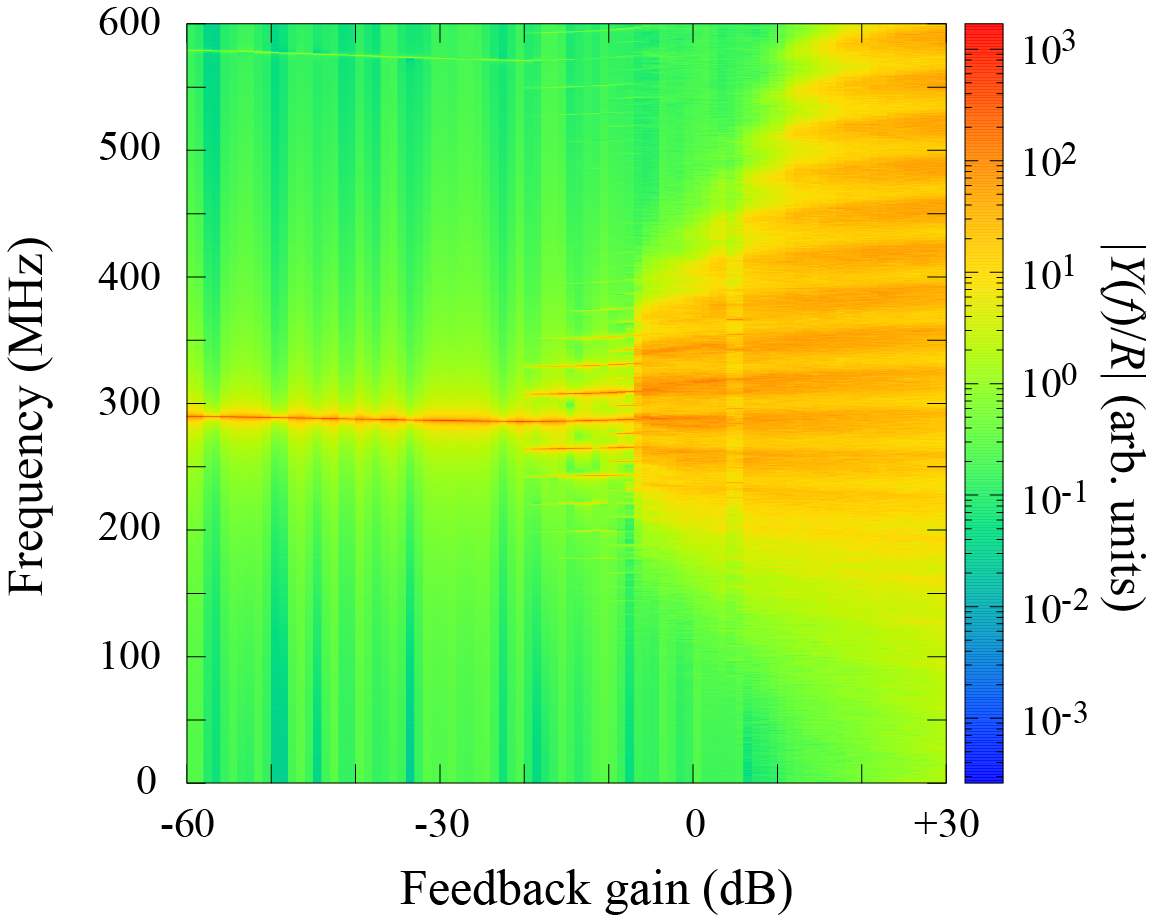}}
\caption{
            Fourier spectra of $Y(t)/R$ as a function of feedback gain. 
         \vspace{-3ex}}
\label{fig:fig2}
\end{figure}


Figure \ref{fig:fig2} shows the Fourier spectra of $Y(t)/R$ obtained by solving Eq. (\ref{eq:Thiele}). 
When the feedback gain is small, a single peak around $290$ MHz is observed. 
Multiple peak structure appears when the feedback gain exceeds roughly over $-20$ dB, which corresponds to feedback magnetic field of $2.2$ Oe. 
As the feedback gain further increases, it becomes difficult to identify the main peak. 


\begin{figure}
\centerline{\includegraphics[width=1.0\columnwidth]{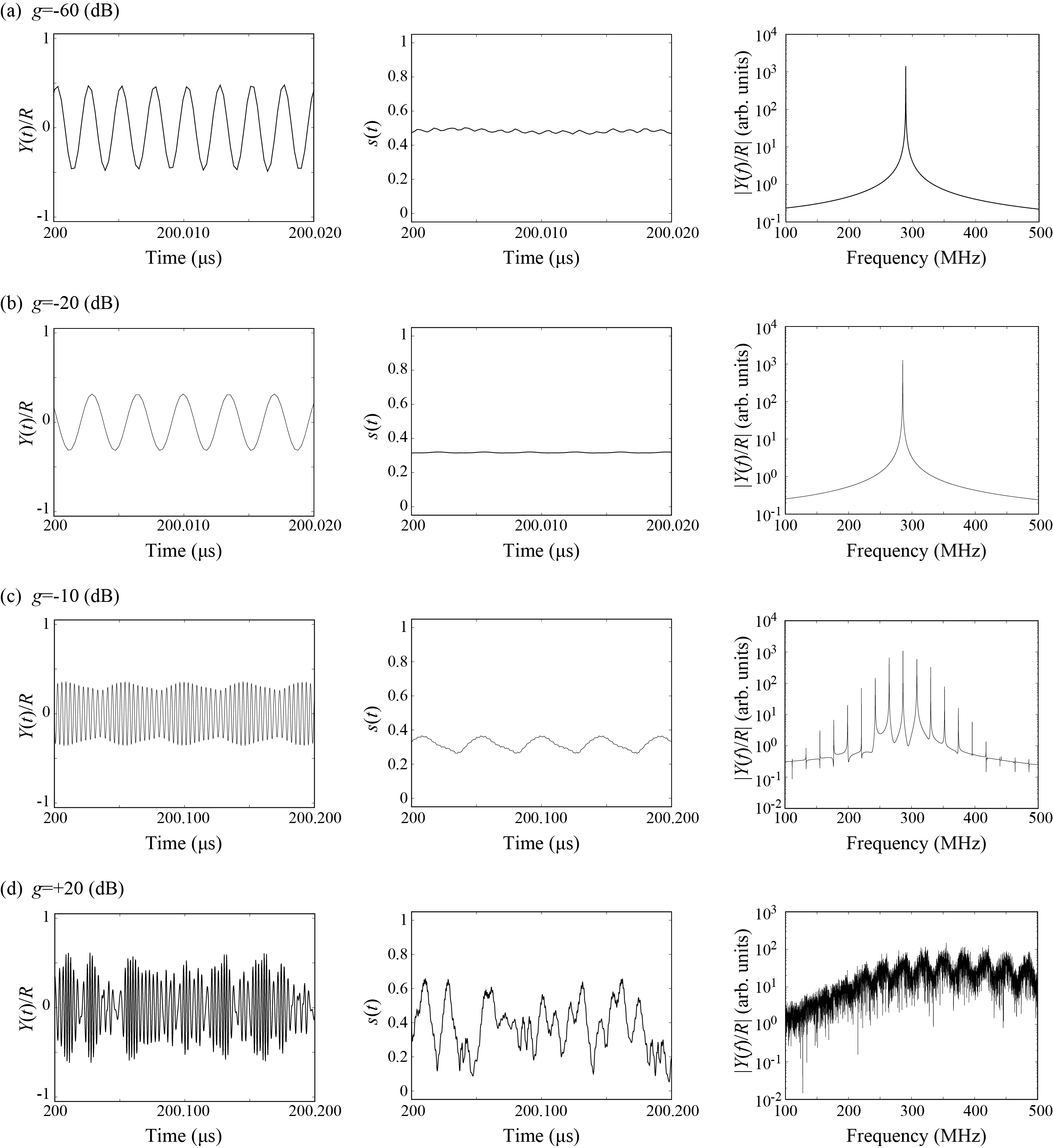}}
\caption{
            (a) Time evolution of $Y(t)/R$ and $s(t)$ in a steady state, and the Fourier transformation $|Y(f)/R|$ of $Y(t)/R$. 
                Feedback gain is $g=-60$ dB, corresponding to $h_{y}=0.1$ Oe. 
            The feedback gain is changed as (b) $g=-20$ dB ($h_{y}=2.2$ Oe), (c) $g=-10$ dB ($h_{y}=4.6$ Oe), and (d) $g=+20$ dB ($h_{y}=46.4$ Oe). 
            Note that the range of time is 20 ns in (a) and (b) while it is $200$ ns in (c) and (d). 
         \vspace{-3ex}}
\label{fig:fig3}
\end{figure}



\begin{figure}
\centerline{\includegraphics[width=0.7\columnwidth]{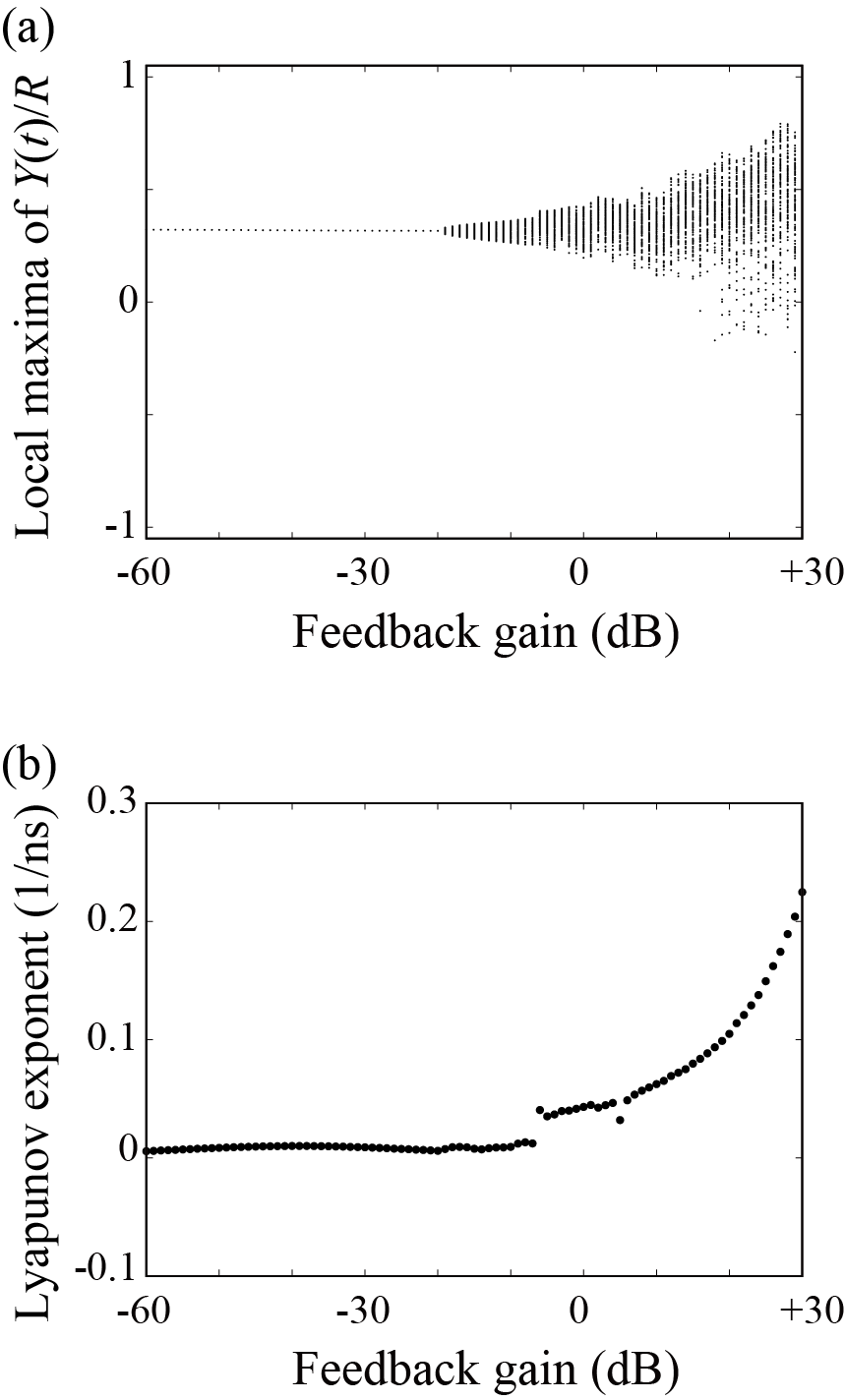}}
\caption{
           (a) Bifurcation diagram summarizing local maxima of $Y(t)/R$ and (b) Lyapunov exponent. 
         \vspace{-3ex}}
\label{fig:fig4}
\end{figure}


The change of the peak structure reflects the temporal dynamics of the vortex core. 
In Fig. \ref{fig:fig3}, we show the temporal dynamics for various gains. 
For example, Fig. \ref{fig:fig3}(a) shows the temporal dynamics of $Y(t)/R$ and $s(t)$, as well as the Fourier spectrum, $|Y(f)|/R$, at the feedback gain of $g=-60$ dB, which corresponds to $h_{y}=0.1$ Oe. 
In this case, the feedback magnetic field is approximately negligible. 
Thus, an auto-oscillation well studied previously \cite{dussaux12,grimaldi14} is excited, where the vortex core rotates around the disk center with practically constant oscillating amplitude (radius). 
Therefore, $Y(t)/R$ is similar to a trigonometric function, and $s(t)$ is approximately constant, as shown in Fig. \ref{fig:fig3}(a).  
A small-amplitude oscillation of $s(t)$ originates from the terms related to $p_{x}$ in Eq. (\ref{eq:Thiele}), as well as the feedback magnetic field, which breaks the axial symmetry around the $z$ axis of Eq. (\ref{eq:Thiele}). 
As can be seen, however, the amplitude of the oscillation in $s(t)$ is very small.  
The Fourier spectrum shows a sharp single peak, as mentioned above. 

When the feedback gain is $-20$ dB ($h_{y}=2.2$ Oe), the vortex core still shows a simple auto-oscillation, where the Fourier spectrum shows a single peak; see Fig. \ref{fig:fig3}(b). 
However, the oscillation amplitude $s(t)$ becomes small compared with that in Fig. \ref{fig:fig3}(a). 
The peak frequency of the Fourier spectrum also becomes slightly low. 
The reason will be explained theoretically in the next section. 

When the feedback gain is further increased to $-10$ dB ($h_{y}=4.6$ Oe), the dynamics become complex, as shown in Fig. \ref{fig:fig3}(c). 
In a short time range, the temporal dynamics of $Y(t)/R$ looks similar to the simple dynamics shown in Figs. \ref{fig:fig3}(a) and \ref{fig:fig3}(b). 
However, when the temporal dynamics over a relative long time range are shown, we find an amplitude modulation of the oscillation. 
This tendency is apparently shown by $s(t)$. 
The Fourier spectrum shows multiple peak structure around a main peak.  

Chaotic dynamics appear clearly when the feedback gain is sufficiently large; see Fig. \ref{fig:fig3}(d), where $g=+20$ dB ($h_{y}=46.4$ Oe). 
The temporal dynamics become non-periodic, and the Fourier spectrum shows a flat structure over a wide range of the frequency. 
In other words, the oscillation frequency of the vortex core is not unique. 
These features are typical in chaotic dynamics \cite{alligood97,ott02}. 

The change of the temporal dynamical behavior can be investigated in a different way \cite{strogatz01}. 
We evaluate a bifurcation diagram, as shown in Fig. \ref{fig:fig4}(a), which summarizes local maxima of $Y(t)/R$. 
When the dynamics is a simple oscillation described by a trigonometric function, for example, the values of the local maxima are always the same, and thus, they are summarized at a single point. 
On the other hand, when the dynamics becomes complex such as chaos, the local maxima show a broadened structure. 
In Fig. \ref{fig:fig4}(a), the local maxima summarize at a single point when the feedback gain is smaller than $20$ dB. 
The local maxima show broadened and approximately symmetric structure when the feedback gain is about in the range of $-20 \lesssim g \lesssim -5$ dB, which is typical in the amplitude-modulation dynamics. 
The maxima are largely broadened and become asymmetric when the feedback gain exceeds about $-5$ dB, which is a typical feature of chaotic dynamics \cite{strogatz01}. 
We also evaluate the Lyapunov exponent, as shown in Fig. \ref{fig:fig4}(b), where the numerical method to evaluate the Lyapunov exponent in a feedback system was developed in Ref. \cite{taniguchi19}. 
The Lyapunov exponent is close to zero for the feedback gain of $g\lesssim -5$ dB, indicating that the dynamics is periodic. 
The Lyapunov exponent becomes largely positive when the feedback gain exceeds about $-5$ dB, which evidently shows the appearance of chaos. 
These results are consistent with the temporal dynamics shown in Fig. \ref{fig:fig3}, where the simple oscillations in Figs. \ref{fig:fig3}(a) and \ref{fig:fig3}(b) and the oscillation with the amplitude modulation in Fig. \ref{fig:fig3}(c) are periodic, while the dynamics shown in Fig. \ref{fig:fig3}(d) are chaos.

Summarizing them, the vortex core dynamics changes from a simple auto-oscillation to complex dynamics such as an oscillation with an amplitude modulation and chaos. 
These changes of the dynamical behavior can be confirmed from both temporal dynamics of the vortex core and its Fourier spectra, which can be experimentally measured due to the magnetoresistance effect.


\section{Theoretical analyses}
\label{sec:Theoretical analyses}


Here, we discuss the bifurcation of the dynamical behavior found in the numerical simulation from theoretical viewpoint. 
First, we should note that it is difficult to obtain exact solutions of Eq. (\ref{eq:Thiele}) analytically, even when the feedback effect is absent. 
However, by applying some approximations, some of the features, such as the appearance of multipeak structure, can be qualitatively explained. 
For this purpose, we express the Thiele equation in terms of $s=\sqrt{X^{2}+Y^{2}}/R$ and $\psi=\tan^{-1}(Y/X)$, which correspond to the distance of the vortex-core position measured from the disk center and the core's phase from the $x$ axis. 
Neglecting small terms in the Thiele equation but keeping terms related to the feedback magnetic field, the approximated equations of motions for $s$ and $\psi$ become 
\begin{widetext}
\begin{equation}
  \dot{s}
  \simeq
  a s
  -
  b s^{3}
  +
  \frac{c\mu^{*}}{GR}
  \sin\psi
  h_{y} 
  s_{\tau} 
  \sin\psi_{\tau}
  -
  \frac{c\mu^{*} |D|}{G^{2}R}
  \cos\psi
  h_{y}
  s_{\tau}
  \sin\psi_{\tau},
  \label{eq:eq_s}
\end{equation}
\begin{equation}
  \dot{\psi}
  \simeq
  \frac{\kappa}{G}
  \left(
    1
    +
    \zeta s^{2}
  \right)
  +
  \frac{c\mu^{*}}{GRs}
  \cos\psi
  h_{y}
  s_{\tau}
  \sin\psi_{\tau}
  +
  \frac{c\mu^{*}|D|}{G^{2}Rs}
  \sin\psi
  h_{y}
  s_{\tau}
  \sin\psi_{\tau}, 
  \label{eq:eq_psi}
\end{equation}
\end{widetext}
where we introduce $a=(|D|\kappa/G^{2})[(J/J_{\rm c})-1]$ and $b=(|D|\kappa/G^{2})(\xi+\zeta)$ with $J_{\rm c}=|D|\kappa/(Ga_{J}p_{z})$, for simplicity. 
We also introduce notations $s_{\tau}=s(t-\tau)$ and $\psi_{\tau}=\psi(t-\tau)$ for delayed variables. 
We used the fact that $G^{2}\gg |D|^{2}$ (or equivalently, $|D|/G\simeq \alpha \ll 1$) to neglect several terms in Eq. (\ref{eq:Thiele}), but, kept terms related to the feedback terms with the coefficient $|D|/G$. 
The reason will be explained below. 
In the following, we first review the steady state solutions of Eqs. (\ref{eq:eq_s}) and (\ref{eq:eq_psi}) and then discuss the effect of the feedback magnetic field


\subsection{Steady state solution in an unperturbed system}
\label{sec:Steady state solution in an unperturbed system}

Note that Eqs. (\ref{eq:eq_s}) and (\ref{eq:eq_psi}) without the feedback effect (i.e., in the limit of $h_{y}\to 0$) are identical to the Stuart-Landau equation \cite{kuramoto03}, which was introduced by Landau to phenomenologically describe dynamics in turbulence and derived by Stuart from hydrodynamic equations. 
The original Stuart-Landau equation is known to be solved analytically \cite{pikovsky03} by introducing a generalized phase $\theta=\psi+[\zeta\kappa/(Gb)]\ln s$.  
Equations (\ref{eq:eq_s}) and (\ref{eq:eq_psi}) in the absence of the feedback effect have the steady state solution given by 
\begin{align}
  \lim_{t\to\infty}
  s
  =
  \sqrt{
    \frac{a}{b}
  }
  \equiv
  s_{0},
&&
  \lim_{t\to\infty}
  \dot{\psi}
  =
  \frac{\kappa}{G}
  \left(
    1
    +
    \zeta
    s_{0}^{2}
  \right), 
\end{align}
when $J/J_{\rm c}>1$, while $\lim_{t\to\infty}s=0$ when $J/J_{\rm c}\le 1$. 
These solutions for $J/J_{\rm c}>1$ correspond to the auto-oscillation of the vortex core with a distance $s_{0}$ from the disk center and the frequency $[\kappa/(2\pi G)](1+\zeta s_{0}^{2})$. 
On the other hand, when $J/J_{\rm c}\le 1$, the vortex core cannot move from the disk center. 
From this perspective, $J_{\rm c}$ can be regarded as a threshold current density for exciting an auto-oscillation. 
For the latter discussion, it is useful to notice that the threshold current density is determined by the parameter $a$, which is a coefficient of a linear term in Eq. (\ref{eq:eq_s}). 
The coefficient $a$ can be rewritten as 
\begin{equation}
  a
  =
  \frac{a_{J}Jp_{z}}{G}
  -
  \frac{|D|\kappa}{G^{2}}, 
  \label{eq:a_def}
\end{equation}
where the first and second terms on the right-hand side correspond to the spin-transfer and damping torques, respectively (recall that $|D|$ is proportional to the damping constant $\alpha$). 
Therefore, $a$ is positive (negative) when the spin-transfer torque is larger (smaller) than the damping torque. 
We can confirm that this condition, $a>(<)0$, is equivalent to $J/J_{\rm c}>(<)1$. 
These characteristics will be used in the following section to study the modulation of the instability threshold by the feedback effects. 
 
  
\subsection{Threshold current and oscillation frequency in the presence of feedback magnetic field}  
\label{sec:Threshold current and oscillation frequency in the presence of feedback magnetic field}  
  
In the presence of the feedback effect, analytical solutions of Eqs. (\ref{eq:eq_s}) and (\ref{eq:eq_psi}) are hardly obtained. 
Recall, however, that the numerical simulation indicated that periodic dynamics can appear even in the presence of the feedback magnetic field when the feedback gain is relatively small. 
Thus, we limit ourselves to a case of a simple auto-oscillation, where $s$ is approximately constant and the vortex-core dynamics are still periodic. 
This approximation is valid when the magnitude of the feedback effect is relatively weak. 
We denote an angular frequency of this oscillation as $\varOmega$, i.e., $\dot{\psi}=\varOmega$. 
Then, we can assume that $\psi=\varOmega t$ and $\psi_{\tau}=\varOmega(t-\tau)$. 
It should be emphasized here that the value of $\varOmega$ is, at this moment, unknown. 
Since $s$ is assumed to be constant, by averaging Eq. (\ref{eq:eq_s}) over a period $2\pi/\varOmega$, the solution of $\dot{s}=0$ after averaging is determined by an equation, 
\begin{equation}
  a
  s
  -
  b
  s^{3}
  +
  \beta 
  s 
  \left(
    \cos\varOmega \tau
    +
    \frac{|D|}{G}
    \sin\varOmega \tau
  \right)
  =
  0,
  \label{eq:steady_state_solution}
\end{equation}
where we introduce a notation $\beta=c\mu^{*}h_{y}/(2GR)$, for simplicity. 
The steady state solution of $s$ in an oscillating state can be obtained by replacing $a$ in the unperturbed solution $s_{0}=\sqrt{a/b}$ with 
\begin{equation}
  a^{\prime}
  =
  a
  +
  \beta 
  \left(
    \cos\varOmega \tau 
    + 
    \frac{|D|}{G}
    \sin\varOmega\tau
  \right).
\end{equation}
Note that the solution $\sqrt{a^{\prime}/b}$ is real number only when $J/J_{\rm c}>1-[G\beta/(|D|\kappa)][\cos\varOmega\tau+(|D|/G)\sin\varOmega\tau]$. 
It means that the feedback magnetic field modulates the threshold current density as 
\begin{equation}
  J_{\rm c}^{\prime}
  =
  J_{\rm c}
  \left[
    1
    -
    \frac{G\beta}{|D|\kappa}
    \left(
      \cos\varOmega \tau
      +
      \frac{|D|}{G}
      \sin\varOmega \tau
    \right)
  \right].
  \label{eq:Jc}
\end{equation}
Here, the following three points should be noticed. 

First, the modulation of the threshold current density comes from the fact that the applied magnetic field originates from the feedback effect. 
This argument can be confirmed from the fact that the last term in Eq. (\ref{eq:steady_state_solution}) includes $s$ coming from the feedback signal, and thus, it contributes to modify $a$ to $a^{\prime}$. 
If the applied magnetic field is, for example, an external oscillating field, which is independent of $s$, such magnetic field does not modify the coefficient $a$ and also does not change the condition of the instability threshold. 
Rather, the phenomenon excited by such magnetic field will be synchronization \cite{pikovsky03}. 

Second, although a modulation of the threshold current density by the feedback effect was found in Ref. \cite{khalsa15}, where the feedback signal was injected as electric current, the way in which the feedback effect modulates the threshold current density is different. 
In the case of the feedback electric current, a term related to the feedback effect appears in the denominator in the definition of the threshold current; see Eq. (6) of Ref. \cite{khalsa15}. 
On the other hand, in the case of the feedback magnetic field studied here, the terms related to the feedback effect appear as a coefficient of proportion, as shown in Eq. (\ref{eq:Jc}). 
The difference comes from the difference of the roles for magnetic field and electric current on an excitation of auto-oscillation. 
The feedback electric current appears as a coefficient multiplied to the current density $J$ in Eq. (\ref{eq:a_def}), and thus, it can be regarded as a modulation of the spin-transfer torque \cite{khalsa15}. 
On the other hand, the feedback magnetic field, which corresponds to term related to $\beta$ in Eq. (\ref{eq:steady_state_solution}), does not include $J$. 
Thus, the term related to the feedback effect of the magnetic field is regarded as an additional to the second term (damping torque) on the right-hand side of Eq. (\ref{eq:a_def}). 
Therefore, the ways the feedback effects modify the threshold current are different for the feedback electric current and magnetic field. 
Note also that the sign of this modulation can be either positive or negative, depending on the values of $\varOmega$ and $\tau$. 

Third but most importantly, at this moment, we cannot estimate the value of Eq. (\ref{eq:Jc}) quantitatively because the value of $\varOmega$ is unknown. 
This is the reason why we kept terms related to $|D|/G$ in Eqs. (\ref{eq:eq_s}) and (\ref{eq:eq_psi}), although $|D|/G\ll 1$; depending on the values of $\varOmega$ and $\tau$, for example, $\cos\varOmega\tau$ in Eq. (\ref{eq:steady_state_solution}) might becomes smaller than $(|D|/G)\sin\varOmega\tau$. 
Therefore, for generality, terms having the factor $|D|/G$ in Eqs. (\ref{eq:eq_s}) and (\ref{eq:eq_psi}) were not neglected. 
The value of $\varOmega$ should be determined from the following equation, which is obtained by averaging Eq. (\ref{eq:eq_psi}) over a period $2\pi/\varOmega$ and using $\dot{\psi}=\varOmega$; 
\begin{equation}
\begin{split}
  \varOmega
  =&
  \frac{\kappa}{G}
  \left(
    1
    +
    \zeta
    \frac{a + \beta \cos\varOmega\tau}{b}
  \right)
\\
  &-
  \beta
  \left(
    \sin\varOmega\tau
    -
    \frac{|D|}{G}
    \cos\varOmega\tau
  \right). 
  \label{eq:angular_frequency}
\end{split}
\end{equation}
Note that a general solution of Eq. (\ref{eq:angular_frequency}) cannot be obtained analytically. 
It becomes possible only when the delay time $\tau$ is sufficiently shorter than $1/\varOmega$ and thus, we can use approximations, $\cos\varOmega\tau\simeq 1$ and $\sin\varOmega\tau\simeq \varOmega\tau$. 
It should also be noted that, to determine the value of $\varOmega$ quantitatively from Eq. (\ref{eq:angular_frequency}), the value of the current density $J$ should be given because it is included in $a$. 
Accordingly, in a strict sense, it is difficult to theoretically predict the threshold current density and the oscillation frequency in the presence of the feedback effect. 
Let us explain this point in detail in the following sections. 


\subsection{Difficulty of the prediction of threshold current density and oscillation analytically}
\label{sec:Difficulty of the prediction of threshold current density and oscillation analytically}

In the absence of the feedback effect, the threshold current density $J_{\rm c}$ can be estimated just from the material parameters and the disk size, and its estimation does not require any information on the oscillation frequency. 
The auto-oscillation is excited when the magnitude of the applied current density exceeds $J_{\rm c}$. 
The oscillation frequency is then given by $[\kappa/(2\pi G)](1+\zeta s_{0}^{2})$, and its value can also be estimated from the material parameters, the disk size, and the current magnitude. 

On the other hand, in the presence of the feedback effect, the threshold current density depends not only on the material parameters and the disk size, in addition to the delay time $\tau$, but also on the frequency $\varOmega/(2\pi)$, as can be seen in Eq. (\ref{eq:Jc}). 
Thus, to estimate the threshold current density, we need to estimate the value of the frequency. 
The frequency is determined by solving Eq. (\ref{eq:angular_frequency}) with respect to $\varOmega$, where an analytical expression of the frequency is hardly obtained, in general.  
Note also that this frequency depends on the current density $J$ through the parameter $a$, and thus, we need to alter some value of $J$ for the estimation of the frequency. 
However, at this moment, it is not clear whether this current density $J$ exceeds the value of $J_{\rm c}^{\prime}$ given by Eq. (\ref{eq:Jc}) because $J_{\rm c}^{\prime}$ depends on $\varOmega$. 
Therefore, even after determining the frequency from Eq. (\ref{eq:angular_frequency}) by substituting the value of $J$, if $J_{\rm c}^{\prime}$ estimated by using this frequency is higher than the current density $J$, an oscillation cannot exist, and the vortex core will be relaxed to the disk center. 

As can be seen in this explanation, in the presence of the feedback effect, it is necessary to solve Eqs. (\ref{eq:Jc}) [or Eq. (\ref{eq:steady_state_solution})] and (\ref{eq:angular_frequency}) simultaneously and self-consistently for the estimation of the threshold current and the oscillation frequency. 
Therefore, the analytical approach to estimate them is very complex. 
If the strength of the feedback effect is sufficiently weak, the frequency $\varOmega/(2\pi)$ might be approximated to the unperturbed one, $[\kappa/(2\pi G)](1+\zeta s_{0}^{2})$, and the threshold current density, Eq. (\ref{eq:Jc}), becomes a periodic function of the delay time $\tau$ \cite{khalsa15}. 
In general, however, the threshold current density and the frequency should be determined simultaneously, as explained above. 


We also explain the reason why the value of $s$ in Fig. \ref{fig:fig3}(b) was smaller than that in Fig. \ref{fig:fig3}(a), although the feedback gain in the former case was stronger than that in the latter case. 
The numerically estimated values of the oscillation frequency, $\varOmega/(2\pi)$, in Figs. \ref{fig:fig3}(a) and \ref{fig:fig3}(b) are $289$ and $285$ MHz, respectively. 
Using these and $\tau=29$ ns, $\cos\varOmega\tau$ is negative for both cases, indicating that the feedback magnetic field increases the magnitude of the threshold current density; see Eq. (\ref{eq:Jc}). 
This increase of the threshold current density is larger for the case in Fig. \ref{fig:fig3}(b) than that in Fig. \ref{fig:fig3}(a) because of the large feedback gain. 
Accordingly, the excitation of the auto-oscillation is easier for the case of the small feedback gain, compared with the case of the large one, and $s$ in Fig. \ref{fig:fig3}(a) becomes larger than that in Fig. \ref{fig:fig3}(b). 
In addition, since the frequency increases as $s$ increases, this difference of $s$ between Fig. \ref{fig:fig3}(a) and \ref{fig:fig3}(b) is consistent with the fact that the oscillation frequency in the former is higher than that in the latter. 
As can be seen in this example, an injection of large feedback effect does not guarantee, for example, a reduction of the threshold current. 
This is because various characteristics, such as the threshold current density and the oscillation frequency, are complex functions of the feedback strength and delay time. 
Nevertheless, as shown here, we can develop the theoretical analyses which qualitatively explain the results of the numerical simulations. 


\begin{figure}
\centerline{\includegraphics[width=1.0\columnwidth]{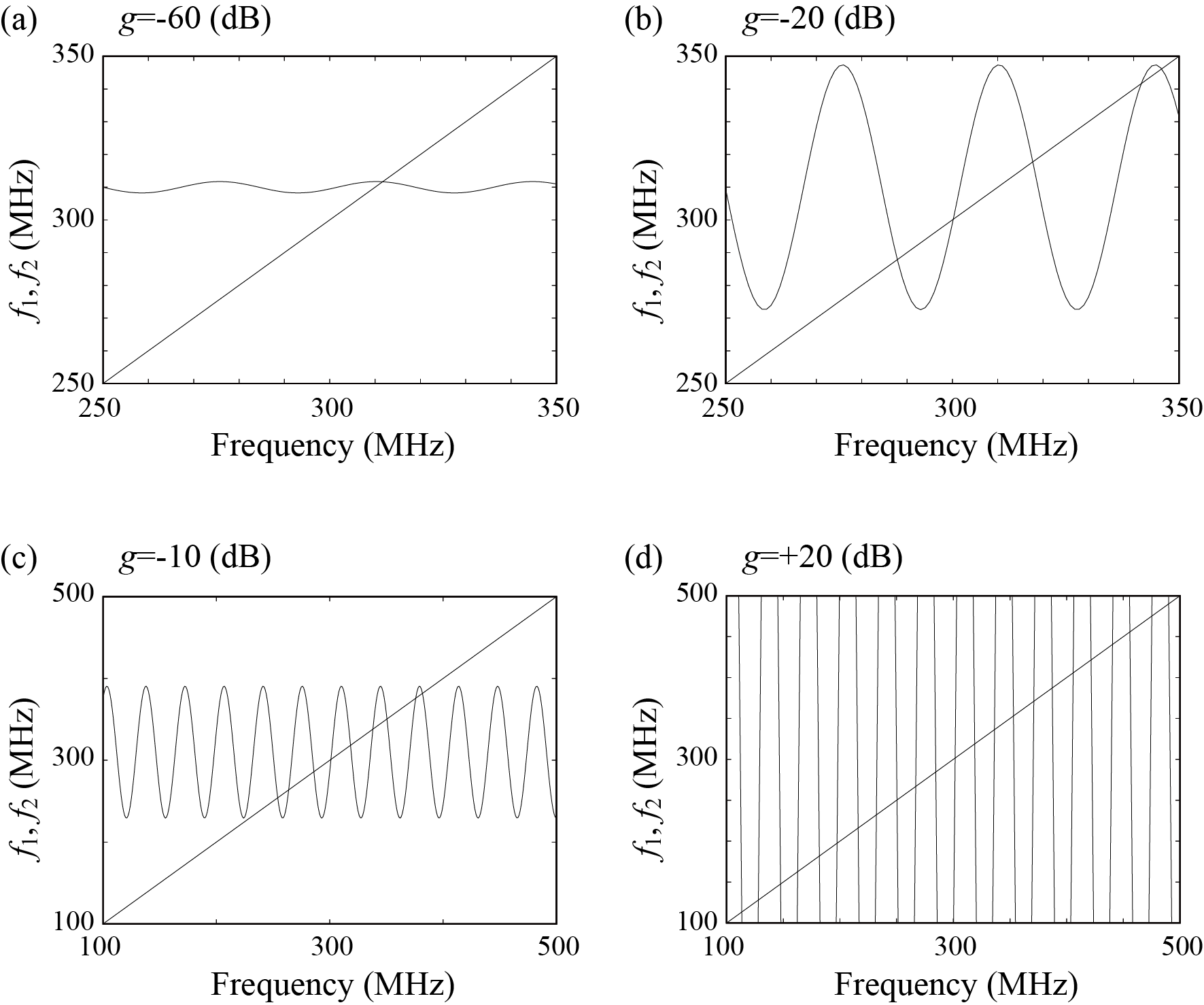}}
\caption{
             Values of $f_{1}$ and $f_{2}$, given by Eqs. (\ref{eq:f1}) and (\ref{eq:f2}), for feedback gains of (a) $g=-60$ dB ($h_{y}=0.1$ Oe),  (b) $g=-20$ dB ($h_{y}=2.2$ Oe), (c) $g=-10$ dB ($h_{y}=4.6$ Oe), and (d) $g=+20$ dB ($h_{y}=46.4$ Oe). 
         \vspace{-3ex}}
\label{fig:fig5}
\end{figure}


\subsection{Theoretical viewpoint on appearance of the multipeak structure}
\label{sec:Theoretical viewpoint on appearance of the multipeak structure}

Finally, we also show that the above theoretical analyses qualitatively explain the appearance of the multipeak structure found in Fig. \ref{fig:fig2}. 
The solution of the frequency satisfying Eq. (\ref{eq:angular_frequency}) can be estimated as intersections of the following two lines; 
\begin{equation}
  f_{1}
  =
  \frac{\varOmega}{2\pi},
  \label{eq:f1}
\end{equation}
\begin{equation}
\begin{split}
  f_{2}
  =&
  \frac{\kappa}{2\pi G}
  \left(
    1
    +
    \zeta
    \frac{a + \beta \cos\varOmega\tau}{b}
  \right)
\\
  &-
  \frac{\beta}{2\pi}
  \left(
    \sin\varOmega\tau
    -
    \frac{|D|}{G}
    \cos\varOmega\tau
  \right). 
  \label{eq:f2}
\end{split}
\end{equation}
Figure \ref{fig:fig5} summarizes examples of $f_{1}$ and $f_{2}$ for some values of the feedback gain. 
For example, in Fig. \ref{fig:fig5}(a), we show $f_{1}$ and $f_{2}$ for the feedback gain of $-60$ dB. 
These two lines intersects at $311$ MHz, which is about $8$\% higher than that estimated from the numerical simulation ($289$ MHz, as mentioned above). 
The difference may come from terms neglected in Eqs. (\ref{eq:eq_s}) and (\ref{eq:eq_psi}). 
Note that there is only one intersection for this case, which agrees with the fact that the Fourier spectrum in Fig. \ref{fig:fig3}(a) shows a single peak. 
When the feedback gain increases, we see a discrepancy between the numerical simulation and the theoretical analyses. 
In Fig. \ref{fig:fig5}(b) for example, we show $f_{1}$ and $f_{2}$ for the feedback gain of $-20$ dB. 
While we find multiple intersections in this figure, the Fourier spectrum in Fig. \ref{fig:fig3}(b) shows a single peak. 
In other words, while the theoretical analyses imply a possibility of an appearance of multiple peak structure, the numerical simulation does not show such behavior. 
The difference again may come from approximations to simplify Eq. (\ref{eq:Thiele}) into the forms of Eqs. (\ref{eq:eq_s}) and (\ref{eq:eq_psi}). 
When the feedback gain further increases, the number of the intersections increases significantly, as can be seen in Figs. \ref{fig:fig5}(c) and \ref{fig:fig5}(d), where the feedback gains are $-10$ and $+20$ dB, respectively. 
In these cases, an assumption that the vortex core oscillates with a unique frequency, i.e., $\dot{\psi}$ is constant, is no longer valid. 
In other words, the appearance of the multi-intersections between $f_{1}$ and $f_{2}$ indicate that the theoretical analyses developed above is no longer applicable, and complex dynamics such as chaos can appear. 
From this perspective, the investigation of the number of the intersections of $f_{1}$ and $f_{2}$ can be useful to estimate qualitatively the feedback gain at which bifurcation from a simple oscillation to complex dynamics occurs.  



\section{Conclusion}
\label{sec:Conclusion}

In summary, the vortex-core dynamics in the presence of the feedback magnetic field were studied by numerical simulation of the Thiele equation. 
The bifurcation from a simple oscillation with a unique frequency to complex dynamics, such as amplitude modulation and chaos, was found from the Fourier spectra, the temporal dynamics, the bifurcation diagram, and the Lyapunov exponent. 
Theoretical analyses on the Thiele equation were also developed to clarify the role of the feedback magnetic field on the vortex-core dynamics. 
The paper evidently shows the difficulty to obtain simple analytical solutions of several characteristics, such as the threshold current density and the oscillation frequency. 
Nevertheless, the theory was able to qualitatively explain several features found in the numerical simulations, such as the change of the oscillation amplitude with changing the feedback gain and the appearance of the multipeak structure. 
The theory also clarifies the difference of the role of the feedback electric current and magnetic field. 


\section*{Acknowledgements}

The work is supported by Japan Society for the Promotion of Science KAKENHI Grants No. 20H05655 and No. 24K01336.


\appendix


\section{Values of parameters}
\label{sec:AppendixA}

The values of the parameters are derived from typical experiments using vortex STOs and feedback circuits \cite{kamimaki21,tsunegi23} and are set as $M=1300$ emu/cm${}^{3}$, $\gamma=1.764\times 10^{7}$ rad/(Oe s), $\alpha=0.010$, $L=5.0$ nm, $R=212.5$ nm, $R_{0}=10$ nm, $P=0.70$, $p_{x}=\sin 60^{\circ}$ ($p_{z}=\cos 60^{\circ}$), $\xi=3.0$, and $\tau=29$ ns. 
The field-like torque is assumed to be zero. 
Using these values, the threshold current density $J_{\rm c}$ in the absence of the feedback effect is estimated as $1.9$ MA/cm${}^{2}$. 
The current density $J$ is defined from the current $I$ as $J=I/(\pi R^{2})$, where the current is $5.5$ mA, corresponding to $J=3.9$ MA/cm${}^{2}$. 
The polarity $p$ and chirality $c$ are each assumed to be $+1$ for convenience. 
The frequency $\kappa/(2\pi G)$ of the ferromagnetic resonance is estimated to be $191$ MHz. 
Some parameters, such as $\zeta$ and $\xi$, might include contributions from Oersted field \cite{dussaux12}. 
Regarding the fact that the current density is constant in this paper, we assume that these contributions are already included in the values shown above. 
In experiments, however, it is difficult to estimate the magnitude of the applied magnetic field directly because it strongly depends on, for example, placement of the metallic line and so on. 
The experimentally controllable parameter is feedback gain (or attenuation), and we use the feedback gain $g$ as a control parameter in the numerical simulation. 
The feedback gain $g$ is defined as $h_{y}=100 \times (10^{g/30}/10)$ Oe, i.e., $g$ is defined so that the feedback gain of $30$ dB corresponds to $h_{y}=100$ Oe.


%



\end{document}